\documentclass[fleqn,11pt]{wlscirep}
\usepackage[utf8]{inputenc}
\usepackage[T1]{fontenc}
\usepackage{braket}
\usepackage{siunitx}
\usepackage{dsfont}
\usepackage{units}
\DeclareSIUnit\angstrom{\text {Å}}
\usepackage{bm}
\renewcommand{\vec}[1]{\boldsymbol{\mathbf{#1}}}

\title{Direct visualization of Rashba-split bands and spin/orbital-charge interconversion at KTaO$_3$ interfaces}
\author[1,$\dagger$]{Sara Varotto}
\author[2,*,$\dagger$]{Annika Johansson}
\author[3,$\dagger$]{Börge Göbel}
\author[1]{Luis M. Vicente-Arche}
\author[1]{Srijani Mallik}
\author[1]{Julien Bréhin}
\author[4]{Raphaël Salazar}
\author[4]{François Bertran}
\author[4]{Patrick Le Fèvre}
\author[5]{Nicolas Bergeal}
\author[4]{Julien Rault}
\author[3]{Ingrid Mertig}
\author[1,*]{Manuel Bibes}

\affil[1]{Unité Mixte de Physique, CNRS, Thales, Université Paris-Saclay, 91767 Palaiseau, France}
\affil[2]{Max Planck Institute of Microstructure Physics, Weinberg 2, 06120 Halle, Germany}
\affil[3]{Institute of Physics, Martin-Luther-Universität Halle-Wittenberg, 06099 Halle, Germany}
\affil[4]{Synchrotron SOLEIL, L’Orme des Merisiers, Saint-Aubin, BP 48, Gif-sur-Yvette Cedex 91192, France}
\affil[5]{Laboratoire de Physique et d’Etude des Matériaux, ESPCI Paris, Université PSL, CNRS, 75005, Paris, France}

\affil[*]{manuel.bibes@cnrs-thales.fr}
\affil[*]{annika.johansson@mpi-halle.mpg.de}

\affil[$\dagger$]{these authors contributed equally to this work}

\begin{abstract}
Rashba interfaces have emerged as promising platforms for spin-charge interconversion through the direct and inverse Edelstein effects. Notably, oxide-based two-dimensional electron gases (2DEGs) display a large and gate-tunable conversion efficiency, as determined by transport measurements. However, a direct visualization of the Rashba-split bands in oxide 2DEGs is lacking, which hampers an advanced understanding of their rich spin-orbit physics. Here, we investigate KTaO$_3$-2DEGs and evidence their Rashba-split bands using angle resolved photoemission spectroscopy. Fitting the bands with a tight-binding Hamiltonian, we extract the effective Rashba coefficient and bring insight into the complex multiorbital nature of the band structure. Our calculations reveal unconventional spin and orbital textures, showing compensation effects from quasi-degenerate band pairs which strongly depend on in-plane anisotropy. We compute the band-resolved spin and orbital Edelstein effects, and predict interconversion efficiencies exceeding those of other oxide 2DEGs. Finally, we suggest design rules for Rashba systems to optimize spin-charge interconversion performance.
\end{abstract}
\begin{document}

\flushbottom
\maketitle

\section*{Introduction}

In its simplest form, the Rashba spin-orbit coupling (RSOC) \cite{bychkov_properties_1984} occurring in systems with broken inversion symmetry lifts the spin degeneracy of a parabolic band pair, by shifting them oppositely along the wave-vector $k$ \cite{manchon_new_2015}. On each of the two corresponding concentric circular Fermi contours, spins are tangential (and thus locked perpendicular to $k$) and curl clockwise for one contour and counter-clockwise for the other. The first direct visualization of Rashba-split bands using angle-resolved photoemission spectroscopy (ARPES) was reported at the surface of heavy metals such as Au \cite{lashell_spin_1996,varykhalov_ir111_2012} or of their alloys \cite{ast_giant_2007} where the RSOC coefficient $\alpha_R$ can take values in the range of a few $\si{\electronvolt \cdot \angstrom}$. The spin-momentum locking of Rashba systems was later harnessed to interconvert spin and charge currents (direct and inverse Rashba-Edelstein effects) at interfaces between heavy metals \cite{rojas-sanchez_spin--charge_2013} and also in oxide-based two-dimensional electron gases (2DEGs) \cite{lesne_highly_2016,vaz2019mapping}. Just a few years after their discovery \cite{ohtomo_high-mobility_2004}, a finite RSOC was indeed identified through weak antilocalization (WAL) in LaAlO$_3$/SrTiO$_3$ (LAO/STO) 2DEGs \cite{caviglia_tunable_2010}, which are non-centrosymmetric systems. $\alpha_R$ was found to amount to a few tens of meV --- considerably lower than at the surface of heavy metals --- and, quite remarkably, to be tuneable by a gate voltage \cite{caviglia_tunable_2010,vaz_determining_2020}. Aside from a report of a giant RSOC at the surface of STO \cite{santander-syro_giant_2014} that later studies did not observe \cite{mckeown_walker_absence_2016}, ARPES could not provide a direct visualization of Rashba-split bands in STO 2DEGs due to the limitation in the energy resolution. This moderate $\alpha_R$ combined with the long scattering time of such 2DEGs was however successfully used to achieve spin-charge interconversion with very high efficiency \cite{lesne_highly_2016,vaz2019mapping}.

Another family of oxide 2DEGs is based on KTaO$_3$ (KTO) instead of STO \cite{gupta_ktao3_2022}. Bulk STO and KTO share several features: they are both quantum paraelectrics and become n-type conductors when doped with minute amounts of impurities \cite{wemple_transport_1965}. Just like STO, KTO may also harbor 2DEGs when interfaced with appropriate materials such as LAO \cite{zhang_unusual_2019}, LaVO$_3$ \cite{wadehra_planar_2020} or reactive metals such as Al (that locally create oxygen vacancies which are n-type dopants) \cite{moreno2021admat}. One major difference though is that Ta is much heavier than Ti and thus RSOC in KTO 2DEGs should be significantly stronger than in STO 2DEGs. Indeed, WAL data yields $\alpha_R \approx$ 300 meV$\cdot$\AA \cite{zhang_unusual_2019} and compatible values were recently derived from bilinear magnetoresistance experiments \cite{moreno2021admat}. Despite several attempts to use ARPES to map the band structure of KTO 2DEGs \cite{santander-syro_orbital_2012,king_subband_2012}, Rashba-split bands were nonetheless never observed, although band structure calculations on KTO \cite{zhang_unusual_2019} and related systems \cite{kim2016strongly} do predict a strong Rashba splitting of the Ta \textit{d}\textsubscript{xz/yz} bands. Therefore, the band structure of (001)-oriented KTO 2DEGs remains elusive to this day, as is the direct visualization of Rashba splitting in any oxide, except for surface states in delafossite single crystals\cite{sunko_maximal_2017}.

In this paper, we report the synthesis of (001)-oriented KTO 2DEGs through the deposition of 1 $\si{\angstrom}$ of Al by molecular beam epitaxy. We use ARPES to measure the band dispersion and associated Fermi surfaces. The data reveal a pair of Rashba-split \textit{d}\textsubscript{xz/yz} bands with a $\alpha_R$ consistent with values extracted from magnetotransport \cite{zhang_unusual_2019,moreno2021admat} and earlier density-functional theory (DFT) calculations \cite{zhang_unusual_2019}. We fit the ARPES data with an 14-band tight-binding (TB) Hamiltonian (see Methods), determine the corresponding spin and orbital textures and compute the Edelstein effect for each band pair. We finally discuss the role of symmetry and anisotropy on the Edelstein effect.

\section*{Results}

\subsection*{Band structure of KTO(001) 2DEGs}

\begin{figure}[ht]
\centering
\includegraphics[width=\linewidth]{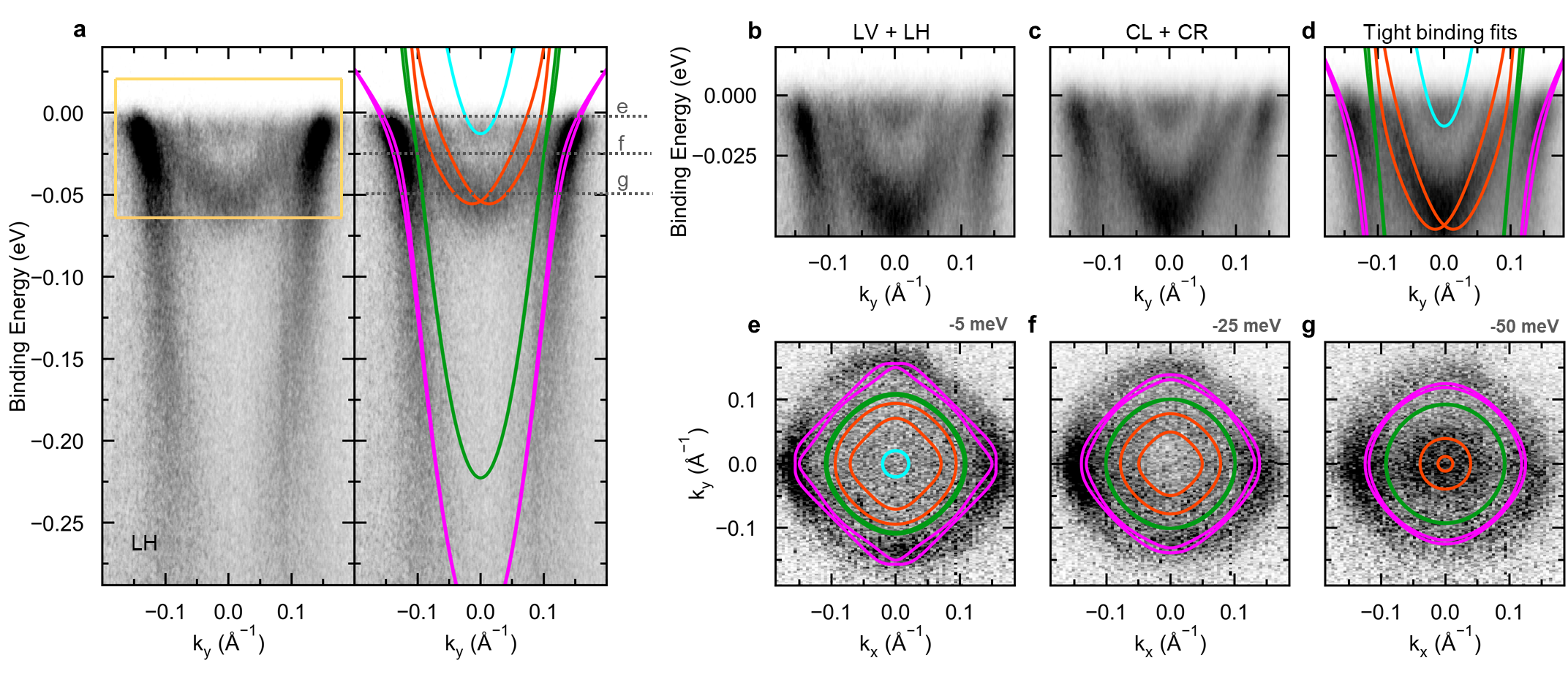}
\caption{\textbf{Electronic band structure of KTO(001) 2DEGs.} \textbf{a} Band dispersion of Al/KTO(001) surface measured by ARPES. The electrons are collected at normal emission with linearly polarized photons at 31 eV ($\Gamma$\textsubscript{002} of bulk KTO). The tight-binding fits are overlaid to the data in the right panel. A specific color is associated to each pair of bands. The yellow box highlights the energy range in panels \textbf{b}-\textbf{d}. \textbf{b} (\textbf{c}) Sum of spectra obtained from two orthogonal linearly (circularly) polarized photons, i.e. linear horizontal and vertical (LV + LH), and circular left and right (CL + CR) respectively. \textbf{d} Same as \textbf{c} with overlaid tight binding fits. Constant energy maps and theoretical contours at \textbf{e} 5 meV, \textbf{f} 25 meV and \textbf{g} 50 meV below the Fermi level. The energies are indicated by gray horizontal lines on the dispersion in panel \textbf{a}.} % Legend (350 words max)
\label{fig:figure1}
\end{figure}

To create an electron gas in a (001)-oriented single crystal of KTO, we deposited 1\textcolor{black}{-2} $\mathrm{\mathring{A}}$ of Al by molecular beam epitaxy (MBE) following the same protocol as detailed in Refs. \cite{rodel_universal_2016,moreno2021admat}. The sample was then transferred in ultra high vacuum to the ARPES measurement chamber (see Methods). Fig.\ref{fig:figure1} displays the detailed band structure of KTO//Al(1 $\mathrm{\mathring{A}}$) near $\Gamma$\textsubscript{002} (corresponding to a photon energy of 31 eV at normal emission). \textcolor{black}{Results from 2 $\mathrm{\mathring{A}}$ samples gave very similar results, albeit with a poorer signal to noise ratio.} We observed a metallic state at the surface, with nearly parabolic bands crossing the Fermi level $\epsilon$\textsubscript{F}. By varying the photon energy we found that the probed states did not significantly disperse with \textit{k}\textsubscript{z}, thereby confirming the quasi-2D nature of the electron gas. In Fig.~\ref{fig:figure1}\textbf{a} we show the band dispersion along the (100) in-plane direction with linear horizontal polarization (LH) of the photon beam. On the right side, the same data are compared with theoretical fits obtained with our TB model. As detailed in the Methods it comprises a total of 14 bands out of which four band pairs fall in the measurement window. The model includes orbital mixing of the 5\textit{d} orbitals of Ta and the strong SOC. For clarity, we associate a specific color to each band pair. Pink, green and orange band pairs result from different linear combinations of the three \textit{t}\textsubscript{2g} orbitals. The first two are predicted to display mainly a \textit{d}\textsubscript{xy} component and a low effective mass \textit{m}\textsubscript{e}= 0.23\textit{m}\textsubscript{0} with \textit{m}\textsubscript{0} the electron mass. We note that although the green bands are not very visible here, they are clearly present on similar data taken at room temperature \cite{moreno2021admat} and in earlier ARPES studies of KTO 2DEGs\cite{santander-syro_orbital_2012,king_subband_2012}, hence we include them in our model. The orange bands instead are mainly formed by mixed \textit{d}\textsubscript{xz} and \textit{d}\textsubscript{yz} orbitals and display a larger effective mass \textit{m}\textsubscript{e}= 0.52\textit{m}\textsubscript{0}. The cyan band pair is introduced in our model as additional \textit{d}\textsubscript{xy} subbands originated from the quantum confinement of the carriers along the \textit{z} direction. However, some uncertainty remains regarding the actual dominant orbital character (i.e. \textit{d}\textsubscript{xy} or \textit{d}\textsubscript{xz, yz}) of this band pair. \textcolor{black}{Please note that in order to describe the Rashba-like band splitting theoretically, we also had to add the \textit{e}\textsubscript{g} states \textit{d}\textsubscript{z\textsuperscript{2}} and the  \textit{d}\textsubscript{x\textsuperscript{2}-y\textsuperscript{2}} to our tight-binding model, even though these states lie above the Fermi energy and are thus not observed in the ARPES measurement. By analogy with Ref. \cite{kim2016strongly}, in Eq. 12 in the Methods section we explain that the orange band pair can be mapped to a Rashba Hamiltonian near the $\Gamma$ point. In this effective 2-band model, the Rashba term is proportional to the orbital mixing coefficient of the \textit{d}\textsubscript{z\textsuperscript{2}} and the \textit{d}\textsubscript{x\textsuperscript{2}-y\textsuperscript{2}} states indicating the necessity to consider the \textit{e}\textsubscript{g} states as well.}

In the measured spectra we can identify two branches close to $\epsilon$\textsubscript{F}, which are nicely fitted with a visible Rashba split band pair (orange). Panels \textbf{b}-\textbf{d} display high resolution measurements to better visualize the splitting of such bands where the binding energy span corresponds to the yellow box in Fig.\ref{fig:figure1}.\textbf{a}. By symmetry, \textit{d}\textsubscript{xz} and \textit{d}\textsubscript{yz} states can be excited only with linear horizontal (LH) and vertical (LV), respectively. Thus, the complete band structure results from the sum of the LV and LH spectra (Fig.~\ref{fig:figure1}.\textbf{b}). Summing circular left and right polarization (CL + CR) yields  a comparable intensity distribution shown in Fig.~\ref{fig:figure1}\textbf{c}, where one can clearly appreciate two \textit{k}-split bands matching with the theoretical ones (Fig.~\ref{fig:figure1}\textbf{d}).   

In Fig.~\ref{fig:figure1}.\textbf{e}, \textbf{f}, \textbf{g} we show constant energy maps in the \textit{k}\textsubscript{x}\textit{k}\textsubscript{y} plane and the corresponding contours derived from our TB fit at binding energies of 5 meV, 25 meV and 50 meV, respectively (gray horizontal lines in panel \textbf{a}). The good agreement between experiment and theory validates the model and the parameters extracted from the fitting of the bands dispersion, e.g. orbital mixing, spin-orbit coupling, hopping amplitudes and on-site energies (discussed in Methods).

\begin{figure}[ht]
\centering
\includegraphics[width=\linewidth]{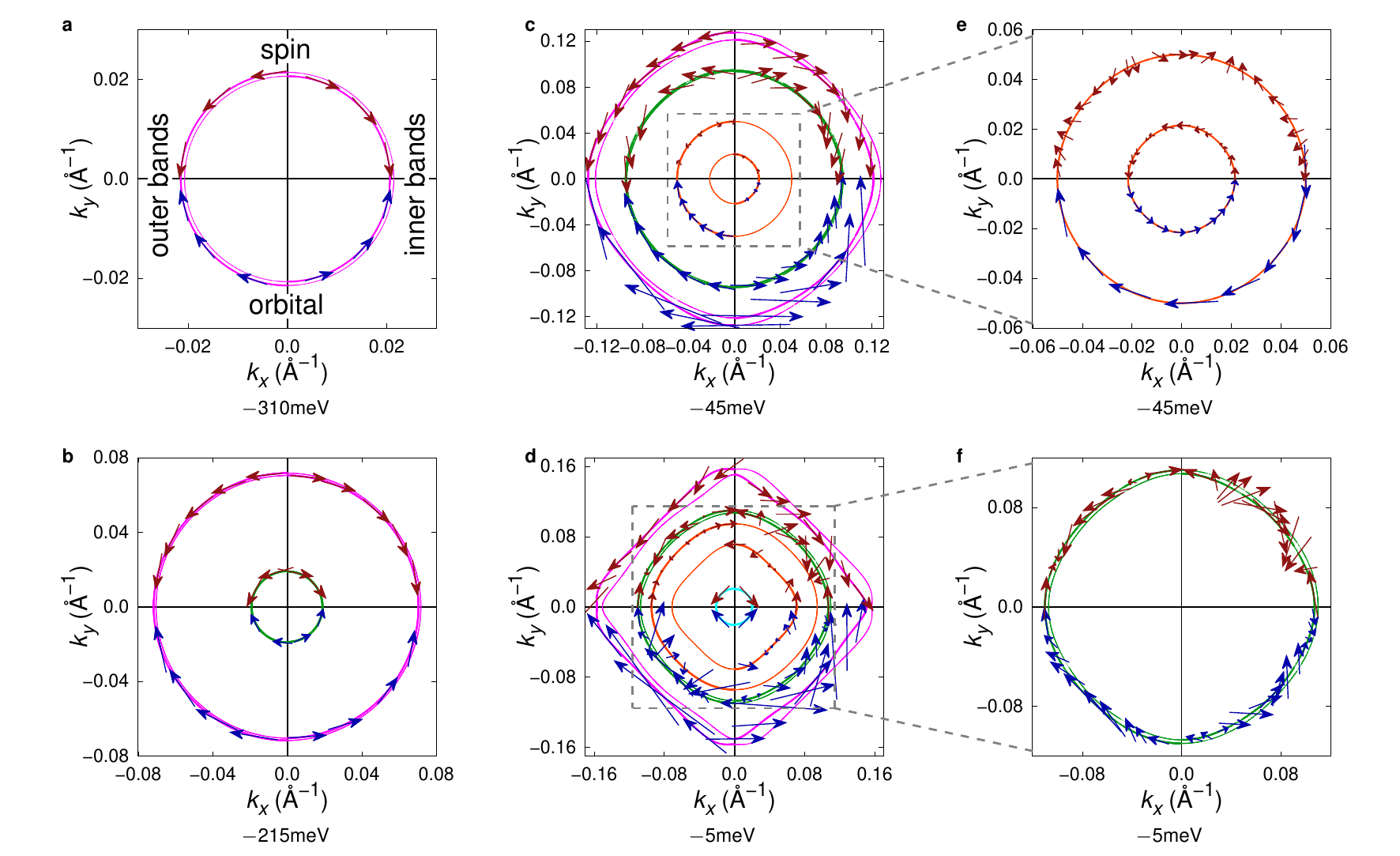}
\caption{\textbf{Spin and orbital textures of electronic states.} Constant energy lines with spin and orbital textures at selected energies. Panels \textbf{a}-\textbf{d}: Iso-energy lines slightly above the band edges of the four band pairs. The upper quadrants display the spin texture (red arrows), the lower quadrants the orbital expectation values (blue arrows). Left quadrants correspond to the outer band of a pair, right quadrants to the inner band. The contours for which the spin or orbital textures are shown are drawn as slightly thicker lines. \textbf{e} detail of panel \textbf{c}. Here, all quadrants show the textures of both bands. \textbf{f} detail of panel \textbf{d}, only the second band pair (green) is shown.} 
\label{fig:figure2}
\end{figure}

\subsection*{Spin and orbital textures}

Next, we analyse the spin and orbital textures of the 2DEG by calculating the expectation values of the spin and orbital moment operators (Eqs.~\eqref{eq:spin_operator} and \eqref{eq:lambda_xyz}) using the eigenstates of the TB Hamiltonian discussed in the Methods section (Eq.~\eqref{eq:Hamiltonian}). Fig.~\ref{fig:figure2} depicts the spin and orbital textures at selected iso-energy lines. Here, we chose energies slightly above the band edge of each band pair. Near the band edge, each band pair has a Rashba-like band structure with almost circular iso-energy lines. The first (pink, Fig.~\ref{fig:figure2}\textbf{a}), second (green, Fig.~\ref{fig:figure2}\textbf{b}), and fourth (cyan, Fig.~\ref{fig:figure2}\textbf{d}) exhibit circular, Rashba-like spin and orbital textures, with the orbital moments pointing in opposite direction to the spin moments. Due to time-reversal and mirror symmetries, both textures are completely in-plane. Near the band edges, the effective Rashba parameters of these  bands, characterizing the band splitting and defined by $\alpha_R = \nicefrac{\Delta k \hbar ^2}{2 m_\mathrm e}$ with $\Delta k$ the difference between the average $k$ of the bands and $\hbar$ the reduced Planck constant, are in the order of 10  $\si{\milli\electronvolt\angstrom}$.
The third band pair (orange) exhibits a much larger splitting, characterized by an effective Rashba parameter of approximately $\alpha_R \approx 320$ $\si{\milli\electronvolt\angstrom}$, which is in good agreement with values deduced from magnetotransport \cite{zhang_unusual_2019,moreno2021admat}. However and quite unexpectedly, the spin and orbital textures deviate from the pure, \textcolor{black}{linear} Rashba model (Fig.~\ref{fig:figure2}\textbf{c} and \textbf{e}). Although the orbital moments show a circular texture, their absolute values at the outer and inner bands differ and are comparably small. The spin texture deviates even more from the standard \textcolor{black}{linear} Rashba model. Spins have comparably small absolute values (up to $0.09 \hbar$ instead of almost $0.5 \hbar$ near the other bands' edges). Further, since the system under consideration deviates from a pure Rashba system, and the corresponding multi-band Hamiltonian also contains higher-order terms in $k$, the spin expectation values at the outer band perform an in-plane rotation of $ 6 \pi$ along the whole iso-energy line, whereas in conventional Rashba systems a $2 \pi$ rotation occurs. Due to the hybridized orbital character of approximately $50\%$ $d_{zx}$ and $50\%$ $d_{yz}$ these strongly Rashba-split bands can approximately be described by a \textcolor{black}{linear} Rashba Hamiltonian in the basis $\{ \frac{1}{\sqrt{2}}(\ket{d_{zx\downarrow}}+i\ket{d_{yz\downarrow}}), \frac{1}{\sqrt{2}}(\ket{d_{zx\uparrow}}-i\ket{d_{yz\uparrow}}) \}$. However, this means that the Rashba-like texture of this band pair occurs in the pseudo spin space $\vec \tau = \vec s_{d_{zx}}-\vec s_{d_{yz}}$, and not in the actual spin space $\vec s = \vec s_{d_{zx}}+\vec s_{d_{yz}}$ which leads to a strong suppression of the spin expectation values~\cite{kim2016strongly} (more details in the Methods Section). Here, $\vec s_{d_{zx}}$ and $\vec s_{d_{yz}}$ are the spins of the corresponding orbitals. Also, since this band pair predominantly is formed by $d_{zx}$ and $d_{yz}$ states, the orbital expectation values are also strongly reduced (the corresponding matrix elements in the $\lambda_x$ and $\lambda_y$ matrices are zero; see Methods section). 

At higher energies, the band structure deviates from the Rashba-like parabolic shape, and the iso-energy lines are no longer circles. Due to hybridization, the spin and orbital textures become more complicated. Most noticeably, their amplitudes vary along an iso-energy line (see e.g. Fig.~\ref{fig:figure2}\textbf{d}), and the amplitude of the orbital moments strongly increases at higher energies. The larger variation in the absolute values of the orbital moments in comparison to the spin expectation values can be explained qualitatively by the higher variety of the values of the corresponding quantum numbers. While there are only two possibilities of projecting the spin to a quantization axis ($s=\frac{1}{2}$), there are five magnetic quantum numbers for the orbital angular momentum of the $d$ electrons under consideration~\cite{johansson2021spin}. This larger range of magnetic quantum numbers is also reflected by stronger varying expectation values of the orbital moment. At approximately $-26$ $\si{\milli\electronvolt}$ the two bands of the second band pair (green) intersect in the $\left\langle10 \right\rangle$ direction, leading to unconventional textures at energies $>-26$ $\si{\milli\electronvolt}$: Fig.~\ref{fig:figure2}\textbf{f} indicates that the spin expectation values rotate by $10 \pi$ along the whole iso-energy line, and the orbital expectation values of the inner band rotate by $6 \pi$ due to deviations of the 2DEG from a pure Rashba system.

\subsection*{Spin and orbital Edelstein effects}

We now discuss the Edelstein effect corresponding to these bands and their spin/orbital textures. We define the Edelstein efficiency tensor $\chi$ by the magnetic moment $\vec m$ per 2D unit cell, induced by the electric field $\vec E$,
\begin{align}\label{eq:Edelstein_efficiency}
    \frac{A_0}{A} \vec m = \chi \vec E = (\chi^\text s + \chi^\text l) \vec E \ .
\end{align}
Here, $A_0$ is the area of the 2D unit cell, $A$ is the area of the sample, and  $\chi^\text s$ and $\chi^\text l$ are the efficiencies of the spin (SEE)\cite{edelstein_spin_1990} and orbital Edelstein effect (OEE)\cite{levitov_magnetoelectric_1985,yoda_current-induced_2015,go_toward_2017,salemi_orbitally_2019}, respectively. The Edelstein efficiency is calculated using the semi-classical Boltzmann approach and a constant relaxation time approximation. For symmetry reasons $\chi_{xy}=-\chi_{yx}$ are the only nonzero tensor elements. Details are discussed in the Methods Section.

\begin{figure}[ht]
\centering
\includegraphics[width=0.5\linewidth]{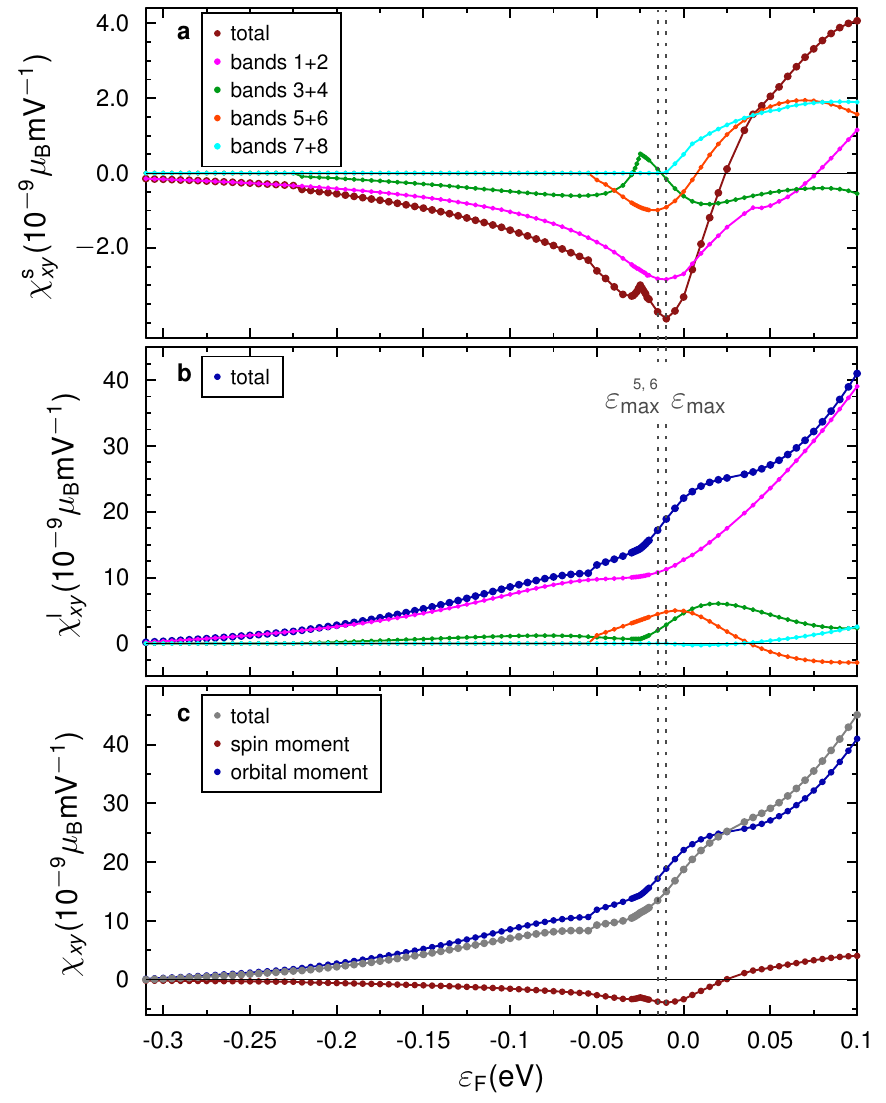}
\caption{\textbf{Edelstein effect.} Edelstein conversion efficiency $\chi_{xy}$ representing the magnetic moment  per surface unit cell along $x$ direction induced by an electric field in $y$ direction. \textbf{a} Contribution from the spin moments (red) and contribution of each band pair (colors as in Fig.~\ref{fig:figure1}). $\epsilon_\text{max}$ corresponds to the energy where the total spin Edelstein efficiency is maximum, $\epsilon_\text{max}^{5,6}$ is the energy of the third band pair's (orange) maximum of the Edelstein efficiency. \textbf{b} Contribution from the orbital moments (blue) and contribution of each band pair. \textbf{c} Total Edelstein efficiency (gray).} 
\label{fig:figure3}
\end{figure}

Fig.~\ref{fig:figure3} depicts the band-resolved and total Edelstein conversion efficiency of the KTO(001) 2DEG versus the Fermi level position. Panels \textbf{a} and \textbf{b} show each band pair's contribution to the SEE and the OEE, respectively, and the total efficiencies as sum of the different pairs (red/blue). Panel \textbf{c} shows the total Edelstein conversion efficiency. Consistent with results on SrTiO$_3$ interfaces~\cite{johansson2021spin}, the OEE dominates the total Edelstein effect, mainly because the orbital moments are in general larger than the spin moments, and the amplitude of the orbital moments of neighboring bands differ more, as discussed above. Usually, the two bands of a pair contribute oppositely to the SEE and OEE, and their contributions partially compensate. If the expectation values differ in magnitude, this compensation is reduced, and the resulting Edelstein effect is enhanced.

When a new band pair is filled, it contributes positively or negatively to the total SEE and OEE, depending on the chirality of the corresponding moments (see Fig.~\ref{fig:figure2}). The fourth band pair (cyan) contributes with a positive sign to the SEE, although it has the same spin chirality as the other bands. Here, the inner band's contribution dominates over the outer band's because of slightly larger spin moments, which inverts the sign of the SEE (see Fig.~\ref{fig:figure2}\textbf{d}).

At $\varepsilon=-26$ $\si{\milli\electronvolt}$, $\chi_{xy}^\text s$ for the second band pair (green) exhibits a sharp kink, which is also visible in the total SEE. This kink is related to the crossing of both bands of this pair, and the related strong modification of the spin texture (see Fig.~\ref{fig:figure2} \textbf{f} and discussion above).

Unexpectedly, the lowest band pair (pink) and not the strongly Rashba-split band pair (orange) dominates the SEE as well as the OEE, although the effective Rashba parameter of the latter exceeds $\alpha_R$ of the lowest band pair by one order of magnitude. This can be directly understood by considering the spin and orbital textures of the strongly Rashba-split band pair (Fig.~\ref{fig:figure2}\textbf{e}). Since the Rashba splitting occurs in the subspace of pseudsopsin $\vec \tau$, and not in the spin subspace (see Methods for details), the spin values are comparably small and deviate from the spin texture of a conventional Rashba system. Due to the orbital character of this band pair (mainly $d_{zx}$ and $d_{yz}$), the orbital texture is small as well. The lowest band pair (pink) has the highest density of states and therefore dominates the total SEE and OEE efficiencies. We note that the maximum spin and orbital EE found here exceed those calculated for STO 2DEGs \cite{johansson2021spin} by a factor of $\sim 2$ and $\sim 4$, respectively.

\section*{Discussion}

As discussed above, the SEE and OEE provided by the third band pair are not as large as one would expect by just considering the giant Rashba-like splitting of this band pair. For symmetry reasons, the states of both bands have equal amount of $d_{yz}$ and $d_{zx}$ character, which leads to an almost complete cancellation of the spin expectation values. 

In order to enhance the corresponding Edelstein efficiency, we reduce the symmetry of the system by introducing an anisotropy with respect to the $d_{yz}$ and $d_{zx}$ orbitals. The simplest way to simulate anisotropy in our particular model Hamiltonian is to assume different on site energies of the corresponding orbitals, $\Delta \epsilon_{yz} \neq \Delta \epsilon_{zx}$. In a realistic system, strain-induced anisotropy would also influence other parameters of the model, but we decided to modify only one parameter in order to demonstrate the general influence of anisotropy, and not to model details of a strained system.

\begin{figure}[ht]
\centering
\includegraphics[width=\linewidth]{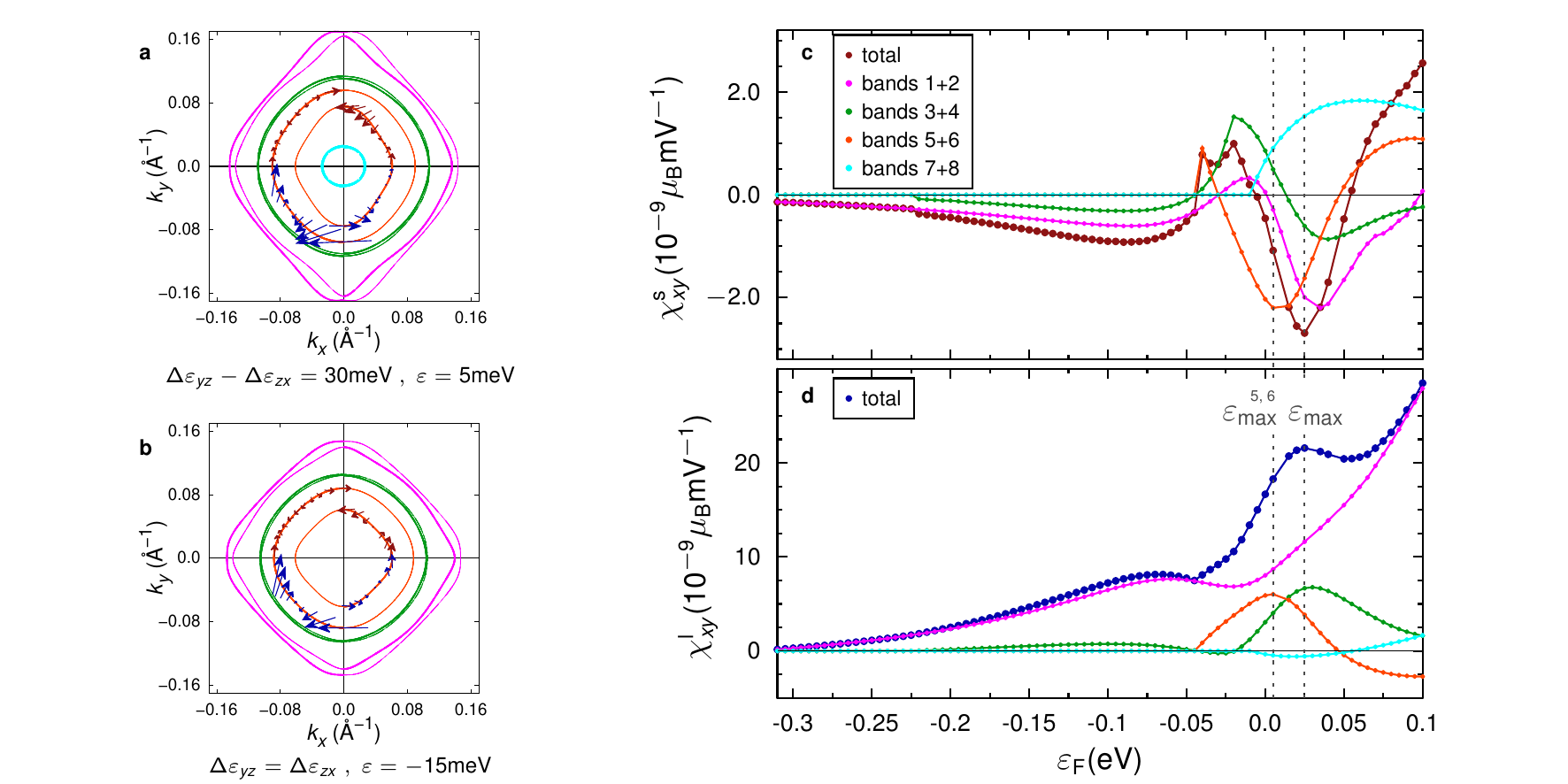}
\caption{\textbf{Influence of anisotropy on the iso-energy lines and the calculated Edelstein efficiency.} \textbf{a} shows the iso-energy lines of a system with $\Delta \epsilon_{yz}-\Delta \epsilon_{zx}=30$ $\si{\milli\electronvolt}$ at $\epsilon= 5 \si {\milli\electronvolt}$, which corresponds to the energy of the maximum contibution of the third band pair (orange) to the spin Edelstein effect. \textbf{b} same as  \textbf{a}, but for the isotropic system. \textbf{c} spin  and \textbf{d} orbital  Edelstein efficiency $\chi_{xy}$ for an anisotropic system with $\Delta \epsilon_{yz}-\Delta \epsilon_{zx}=30 \si{\milli\electronvolt}$.} 
\label{fig:figure4}
\end{figure}

Fig.~\ref{fig:figure4}\textbf{a} and \textbf{b} show the spin and orbital textures of an anisotropic ($\Delta \epsilon_{yz}-\Delta \epsilon_{zx}= 30 \si{\milli\electronvolt}$) and the isotropic 2DEG, respectively. Here, we chose the energies of the third band pair's maximum contribution to the SEE (labeled $\epsilon_\text{max}^{5,6}$). In comparison to the isotropic system, the spin and orbital expectation values of the anisotropic system are remarkably increased, as envisaged from the low-energy expansion of the third band pair (Eq.~\eqref{eq:expansion}). While the character of these bands at the band edge is approximately $50 \%$ of each $d_{yz}$ and $d_{zx}$ in the isotropic case, this ratio strongly changes with anisotropy ($62 \%$ of $d_{zx}$  and $35 \%$ of $d_{yz}$ for $\Delta \epsilon_{yz}-\Delta \epsilon_{zx}= 30 \si{\milli\electronvolt}$). Hence, the $\vec k$ dependent spin expectation values are increased, because the contributions from both orbitals compensate less in the anisotropic case. Further, the admixture of $d_{xy}$ states ($2\%$ at the band edges versus $0.2\%$ in the isotropic case) leads to enhanced orbital expectation values.

Fig. \ref{fig:figure4}\textbf{c} shows the SEE and OEE efficiencies versus the Fermi level for an anisotropic system with $\Delta \epsilon_{yz}-\Delta \epsilon_{zx}= 30 \si{\milli\electronvolt}$. In comparison to the SEE and OEE in the isotropic case (Fig.~\ref{fig:figure3}), the contribution of the third band pair (orange) to the total SEE and OEE signal is indeed increased by a factor $\sim$3 due to the enhanced $\vec k$ dependent spin and orbital expectation values. However, the anisotropy also affects the other band pairs, and the total SEE and OEE, which are superpositions of all bands' contributions, are reduced. Obviously, the Edelstein tensor becomes anisotropic in the presence of strain, $\chi_{xy} \neq - \chi_{yx}$, and in general the magnitude of $\chi_{xy}$ and $\chi_{yx}$ are affected oppositely by anisotropy.

Our work shows that even when the host material contains a heavy element such as Ta, in 2DEGs based on $d$ element perovskite oxides and accordingly displaying a substantial Rashba splitting, the orbital symmetry may lead to strong compensation effects, limiting the spin-charge interconversion efficiency through the EE and inverse EE. The compensation effects are three-fold: (i) the EE from the $d_{xz/yz}$ band is weaker than expected due to the opposite spin textures of $d_{xz}$ and $d_{yz}$ components; (ii) the EE from the $d_{xy}$ bands and $d_{xz/yz}$ bands typically have opposite sign; (iii) the spin EE and orbital EE generally have opposite signs, the total response being dominated by the orbital EE. 

Our calculations indicate that introducing in-plane anisotropy is necessary to unleash the potential of the $d_{xz/yz}$ Rashba split bands for spin-charge interconversion in KTO 2DEGs. This can be achieved from a cubic material like KTO by making it orthorhombic, e.g. by growing it on orthorhombic substrates such as rare-earth scandates which have a good lattice match with KTO. Another approach to lift the $d_{xz/yz}$ orbital degeneracy is to design a 2DEG from a material that, unlike KTO, is orthorhombic in the bulk. Candidates include SrZrO$_3$, CaZrO$_3$ or CaHfO$_3$. DFT simulations predict that this latter compound, that includes the 5$d$ element Hf, can be made metallic through doping with oxygen vacancies \cite{alay-e-abbas_chemical_2014}, similar to STO and KTO, which suggests that it may host a 2DEG when interfaced with another oxide. 

To suppress the second compensation effect, it may be attractive to generate 2DEGs in oxide materials in which crystal field splitting causes the $d_{xz/yz}$ bands to lie below the $d_{xy}$ bands. Recent ARPES measurements suggested that such a band order inversion occurs at interfaces between STO and $\gamma$-Al$_2$O$_3$ \cite{chikina_band-order_2021}. It would be 
interesting to see if the same inversion could be engineered in KTO 2DEGs. We note that in anatase TiO$_2$, which was shown to be a suitable 2DEG host\cite{sarkar_electron_2015}, the $d_{xz/yz}$ bands also lie below the $d_{xy}$ bands. This material and related members of its family, if properly engineered, thus look promising for efficient spin-charge interconversion by EE and inverse EE.

Finally, although this is not true at all energies, in KTO 2DEGs the spin EE and the orbital EE typically compete with each other. In classical spin pumping experiments only a spin current is injected and the conversion to a charge current is mostly driven by the spin EE \cite{vaz2019mapping}. However, in more advanced schemes aiming to harness orbital-charge interconversion as well \cite{go_orbitronics_2021} it would be beneficial to work with systems in which the spin EE and orbital EE have the same sign. Materials with more than half-filled orbitals, in which the spin and orbital moments should be parallel, could represent a fruitful research direction.

\section*{Methods}

\textcolor{black}{
\subsection*{Sample preparation}
After preannealing the KTO at 300$^\circ$C for 1 h in UHV, we grew 1-2 $\si{\angstrom}$ of Al at room temperature at 7$\times$10$^{-10}$ mbar using a Knudsen cell heated to 1000$^\circ$C at a growth rate of 0.011 $\si{\angstrom}$.s$^{-1}$. The samples were then transferred in UHV to the connected ARPES chamber. Low-energy electron diffraction performed on the samples  before and after Al deposition showed sharp diffraction spots corresponding to a square lattice, attesting to the high structural coherence of the surface. In situ X-ray photoelectron spectroscopy showed that Al deposition onto KTO led to the reduction of the Ta valence from the nominal 5+, consistent with 2DEG formation \cite{moreno2021admat}, and that Al was fully oxidized to Al$^{3+}$.}

\subsection*{Angular resolved photoemission spectroscopy}
High resolution angular resolved photoemission spectroscopy (ARPES) spectra were collected at the Cassiopée beamline of Synchrotron SOLEIL (France), by a Scienta R4000 electron energy analyser. The beamline allows to control the energy and polarization of the VUV photons. Data presented in the manuscript were collected at different polarization to probe electron states with different orbital symmetries: linear horizontal (i.e. parallel) or vertical to the scattering plane (LH or LV), and circular left or right (CL or CR). 
Considering the sample surface plane \textit{xy} and vertical slit along \textit{y} (LV), for a normal emission geometry the scattering plane is defined as the mirror plane \textit{xz}. This allows to probe only odd symmetry orbitals with respect to \textit{xz} plane, corresponding to \textit{d}\textsubscript{xy} and \textit{d}\textsubscript{yz} orbitals. By using horizontal slit (LH), only \textit{d}\textsubscript{xz} orbitals are selected. The sum of the LH and LV spectra would thus contain contribution from all the \textit{e}\textsubscript{g} bands. 
The sample was kept at 15 K by liquid He in order to minimize the thermal noise. The energy and angular resolution were 15 meV and <0.25$^\circ$.

\subsection*{Tight-binding model}

We diagonalize the tig
ht-binding Hamiltonian
\begin{align}\label{eq:Hamiltonian}
H=H_\mathrm{hop}+H_\mathrm{SOC}+H_\mathrm{mix}
\end{align}
to describe the electrons gas that is confined at the interface. In the model, the electron gas consists of a single layer square lattice (lattice constant $a=4.0\,\mathrm{\mathring{A}}$) formed by the Ta atoms. Only the $5d$ electrons of the Ta atoms have been considered, since they form the states close to the Fermi energy. The model is similar to our previous works on STO 2DEGs~\cite{vaz2019mapping,johansson2021spin} but instead of considering only the three $t_{2g}$ orbitals, we also consider the $d_{z^2}$ and $d_{x^2-y^2}$ orbitals. For the $d_{xy}$ orbital we consider two additional subbands. In total, the Hamiltonian is a $14\times 14$ matrix which results in 14 bands in the band structure. The basis is 
\begin{align}\label{eq:basis}
\{ \ket{d_{z^2\uparrow}},\ket{d_{z^2\downarrow}},\ket{d_{x^2-y^2\uparrow}},\ket{d_{x^2-y^2\downarrow}},\ket{d_{xy1\uparrow}},\ket{d_{xy1\downarrow}} ,\ket{d_{xy2\uparrow}},\ket{d_{xy2\downarrow}} ,\ket{d_{xy3\uparrow}},\ket{d_{xy3\downarrow}},\ket{d_{zx\uparrow}},\ket{d_{zx\downarrow}},\ket{d_{yz\uparrow}},\ket{d_{yz\downarrow}} \}.
\end{align}

Since we consider a two-dimensional system with hoppings along $x$ and $y$, the hopping matrix $H_\mathrm{hop}$ is diagonal except for a mixing of $d_{z^2}$ and $d_{x^2-y^2}$ states
\begin{align}
H_\mathrm{hop}=\begin{pmatrix}
\epsilon_{z^2}&m_{z^2,x^2-y^2}&0&0&0&0&0\\
m_{z^2,x^2-y^2}&\epsilon_{x^2-y^2}&0&0&0&0&0\\
0&0&\epsilon_{xy1}&0&0&0&0\\
0&0&0&\epsilon_{xy2}&0&0&0\\
0&0&0&0&\epsilon_{xy3}&0&0\\
0&0&0&0&0&\epsilon_{zx}&0\\
0&0&0&0&0&0&\epsilon_{yz}\\
\end{pmatrix}
\bigotimes
\begin{pmatrix} 1&0\\0&1 \end{pmatrix}
\end{align}
%
%Here, $\bigotimes$ multiplies every element of the left matrix with the right matrix to account for the electron spin. 
The elements of this matrix have been determined by using the Slater-Koster formalism 
\begin{align}
\epsilon_{z^2}&=\frac{t_\sigma+3t_\delta}{2}\left[\cos(ak_x)+\cos(ak_y)\right]+\Delta\epsilon_{z^2}\\
\epsilon_{x^2-y^2}&=\frac{3t_\sigma+t_\delta}{2}\left[\cos(ak_x)+\cos(ak_y)\right]+\Delta\epsilon_{x^2-y^2}\\
\epsilon_{xy\{1,2,3\}}&=2t_\pi\cos(ak_x)+2t_\pi\cos(ak_y)+\Delta\epsilon_{xy\{1,2,3\}}\\
\epsilon_{zx}&=2t_\pi\cos(ak_x)+2t_\delta\cos(ak_y)+\Delta\epsilon_{zx}\\
\epsilon_{yz}&=2t_\delta\cos(ak_x)+2t_\pi\cos(ak_y)+\Delta\epsilon_{yz}\\
m_{z^2,x^2-y^2}&=\frac{\sqrt{3}}{2}(t_\delta-t_\sigma)\left[\cos(ak_x)-\cos(ak_y)\right]
\end{align}
Independent hopping amplitudes are $t_\sigma=-0.46\,\mathrm{eV}$, $t_\pi=-1.37\,\mathrm{eV}$ and $t_\delta=-0.07\,\mathrm{eV}$. To account for the band shift due to the broken inversion symmetry at the interface, we take into account the onsite-energies $\Delta\epsilon_{z^2}=3.670\,\mathrm{eV}$, $\Delta\epsilon_{x^2-y^2}=21.45\,\mathrm{eV}$, $\Delta\epsilon_{xy1}=5.180\,\mathrm{eV}$, $\Delta\epsilon_{xy2}=5.275\,\mathrm{eV}$, $\Delta\epsilon_{xy3}=5.515\,\mathrm{eV}$, $\Delta\epsilon_{zx}=2.885\,\mathrm{eV}$ and $\Delta\epsilon_{yz}=2.885\,\mathrm{eV}$. 

The matrix $H_\mathrm{SOC}$ with $\lambda=0.16\,\mathrm{eV}$ describes on-site spin-orbit coupling and mixes different spins and even different orbital
\setcounter{MaxMatrixCols}{14}
\begin{align}
H_\mathrm{SOC}=\frac{2}{3}\lambda\begin{pmatrix}
0&0&0&0&0&0&0&0&0&0&0&-\frac{\sqrt{3}}{2}&0&i\frac{\sqrt{3}}{2}\\
0&0&0&0&0&0&0&0&0&0&\frac{\sqrt{3}}{2}&0&i\frac{\sqrt{3}}{2}&0\\
0&0&0&0&-i&0&-i&0&-i&0&0&\frac{1}{2}&0&\frac{i}{2}\\
0&0&0&0&0&i&0&i&0&i&-\frac{1}{2}&0&\frac{i}{2}&0\\
0&0&i&0&0&0&0&0&0&0&0&-\frac{i}{2}&0&\frac{1}{2}\\
0&0&0&-i&0&0&0&0&0&0&-\frac{i}{2}&0&-\frac{1}{2}&0\\
0&0&i&0&0&0&0&0&0&0&0&-\frac{i}{2}&0&\frac{1}{2}\\
0&0&0&-i&0&0&0&0&0&0&-\frac{i}{2}&0&-\frac{1}{2}&0\\
0&0&i&0&0&0&0&0&0&0&0&-\frac{i}{2}&0&\frac{1}{2}\\
0&0&0&-i&0&0&0&0&0&0&-\frac{i}{2}&0&-\frac{1}{2}&0\\
0&\frac{\sqrt{3}}{2}&0&-\frac{1}{2}&0&\frac{i}{2}&0&\frac{i}{2}&0&\frac{i}{2}&0&0&-\frac{i}{2}&0\\
-\frac{\sqrt{3}}{2}&0&\frac{1}{2}&0&\frac{i}{2}&0&\frac{i}{2}&0&\frac{i}{2}&0&0&0&0&\frac{i}{2}\\
0&-i\frac{\sqrt{3}}{2}&0&-\frac{i}{2}&0&-\frac{1}{2}&0&-\frac{1}{2}&0&-\frac{1}{2}&\frac{i}{2}&0&0&0\\
-i\frac{\sqrt{3}}{2}&0&-\frac{i}{2}&0&\frac{1}{2}&0&\frac{1}{2}&0&\frac{1}{2}&0&0&-\frac{i}{2}&0&0
\end{pmatrix}
\end{align}

Due to the gradient potential at the interface, the oxygen $p$ orbitals are displaced away from the bond connecting two Ta atoms. This allows for hopping terms that are forbidden in the bulk. The effective hopping amplitude in a hopping network between two neighboring Ta $d$ orbitals, via an intermediate hopping to a oxygen $p$ orbital, is finite and antisymmetric~\cite{khalsa2013theory,zhong2013theory} when mixing $d_{zx}$ or $d_{yz}$ with $d_{xy}$, $d_{z^2}$ or $d_{x^2-y^2}$ orbitals. This gives rise to~\cite{kim2016strongly}
\begin{align*}
H_\mathrm{mix}=&2i
\begin{pmatrix}
0&0&0&0&0&-g_2\sin(ak_x)&-g_2\sin(ak_y)\\
0&0&0&0&0&-g_3\sin(ak_x)&g_3\sin(ak_y)\\
0&0&0&0&0&g_{1}\sin(ak_y)&g_{1}\sin(ak_x)\\
0&0&0&0&0&g_{1}\sin(ak_y)&g_{1}\sin(ak_x)\\
0&0&0&0&0&g_{1}\sin(ak_y)&g_{1}\sin(ak_x)\\
g_2\sin(ak_x)&g_3\sin(ak_x)&-g_{1}\sin(ak_y)&-g_{1}\sin(ak_y)&-g_{1}\sin(ak_y)&0&0\\
g_2\sin(ak_y)&-g_3\sin(ak_y)&-g_{1}\sin(ak_x)&-g_{1}\sin(ak_x)&-g_{1}\sin(ak_x)&0&0
\end{pmatrix}\\
&\bigotimes
\begin{pmatrix} 1&0\\0&1 \end{pmatrix}
\end{align*}
with the amplitudes $g_{1}=0.005\,\mathrm{eV}$, $g_{2}=0.5\,\mathrm{eV}$ and $g_{3}=0.002\,\mathrm{eV}$.

Note that we did not observe the $d_{x^2-y^2}$ and $d_{z^2}$ bands in the ARPES measurements, as they are several eV above the Fermi energy. On the one hand, considering the $d_{x^2-y^2}$ orbital was not useful for improving the fit which is why we practically disregarded this band pair by using a large onsite energy $\Delta\epsilon_{x^2-y^2}$. Still, we wanted to include this orbital for the sake of completeness. On the other hand, considering  $d_{z^2}$ improved the fit significantly. In fact, without this orbital we were not able to reproduce the large Rashba splitting of the lower $d_{zx/yz}$ bands. The origin of this effect has been explained in Ref.~\cite{kim2016strongly} for a monolayer of BaHfO$_3$: Near $\Gamma$ the Hamiltonian can be expanded in $\vec{k}$. For the band pair with the strong Rashba splitting (the lower $d_{zx/yz}$ bands), we get
\begin{align}\label{eq:expansion}
H_\mathrm{eff}=
h(\vec{k})
\begin{pmatrix}1&0\\0&1\end{pmatrix}+
\frac{2\sqrt{3}g_2\lambda}{\epsilon_{z^2}-\epsilon_{yz/zx}}\left(\vec{\tau}\times\vec{k}\right)\cdot\vec{e}_z,
\end{align}
if we treat these two bands individually in the basis $\{ \frac{1}{\sqrt{2}}(\ket{d_{zx\downarrow}}+i\ket{d_{yz\downarrow}}), \frac{1}{\sqrt{2}}(\ket{d_{zx\uparrow}}-i\ket{d_{yz\uparrow}}) \}$.
The second term describes the Rashba splitting. It is quantified by $g_2$ (the orbital mixing amplitude of $d_{z^2}$ with $d_{yz}$ and $d_{zx}$), as well as by $\epsilon_{z^2}-\epsilon_{yz/zx}$ (the energy difference of the $d_{yz}/d_{zx}$ band pair and the $d_{z^2}$ band at the $\Gamma$ point). $\vec{\tau}$ describes the pseudo spin $\vec{\tau}=\vec{s}_{d_{zx}}-\vec{s}_{d_{yz}}$ which forms a Rashba spin texture on the Fermi line. However, the actual spin $\vec{s}=\vec{s}_{d_{zx}}+\vec{s}_{d_{yz}}$ is compensated. This explains why the spin texture of the band pair with the strong Rashba splitting is so small in the full model.

\subsection*{Calculation of the spin and orbital Edelstein effect}
We use the semiclassical Boltzman transport theory to calculate the spin and orbital Edelstein efficiencies defined by Eq.~\eqref{eq:Edelstein_efficiency}. The magnetization $\vec m_{\text{s}/\text{l}}$ originating from the spin and orbital moments, resepectively, and induced by the external electric field $\vec E$ is calculated as~\cite{johansson2021spin}
\begin{align}\label{eq:magnetization}
    \vec m_{\text s/\text l} = - \frac{g_{\text s / \text l} \mu_\text B}{\hbar} \sum \limits_{\vec k} f_{\vec k} \Braket{\vec{s}/\vec{ l}}_{\vec k} \ .
\end{align}
Here, $g_{\text s/\text l}$ are the spin and orbital Land{\'e}'s $g$ factors, respectively, which we have set to $g_\text s = 2$ and $g_\text l = 1$ in our calculations.  $\mu_\text B $ is the Bohr magneton, $f_{\vec k}$ is the distribution function, and $\Braket{\vec{s}/\vec{l}}_{\vec{k}}$ is the $\vec k$ dependent expectation value of the spin and orbital moment, respectively, which is calculated by
\begin{align}\label{eq:sl_k}
\Braket{\vec s/ \vec l}_{\vec k} = \Braket{\Psi_{\vec k} | \vec{s}/\vec{l} | \Psi_{\vec{k}}} \ .
\end{align}
$\Ket{\Psi_{\vec k}}$ are the eigenstates of the Hamiltonian~\eqref{eq:Hamiltonian}.
For reasons of clearness, we have merged  the crystal momentum $\hbar \vec{k}$ and the band index $n$ to the multi-index $\vec{k}$ here and in the following. The operators of spin and orbital moment in the basis~\eqref{eq:basis} of the tight-binding Hamiltonian~\eqref{eq:Hamiltonian} are
\begin{align}\label{eq:spin_operator}
    \vec s = \frac{\hbar}{2} \mathds{1}_{7} \otimes \vec \sigma , \ \ \ \ 
    \vec l = \hbar \bm \lambda \otimes \mathds{1}_2 
\end{align}
with 
\begin{align}\label{eq:lambda_xyz}
\begin{split}
  & \lambda_x = \begin{pmatrix}
   0 & 0 & 0 & 0 & 0 & 0 &  \sqrt{3} \ i\\
   0 & 0 & 0 & 0 & 0 & 0 &  i \\
   0 & 0 & 0 & 0 &0 & -  i & 0 \\
   0 & 0 & 0 & 0 &0 & -  i & 0 \\
   0 & 0 & 0 & 0 &0 & -  i & 0 \\
   0 & 0 &  i &  i &  i & 0 & 0 \\
   -  \sqrt{3}  i& -  i & 0 & 0 & 0 & 0 & 0 
   \end{pmatrix}  \  , \ \  \lambda_y = \begin{pmatrix}
   0 & 0 & 0 & 0 & 0 & - \sqrt{3}  i & 0  \\
   0 & 0 & 0 & 0 & 0 &  i & 0  \\
   0 & 0 & 0 & 0 & 0 & 0 &  i \\
  0 & 0 & 0 & 0 & 0 & 0 &  i \\
 0 & 0 & 0 & 0 & 0 & 0 &  i \\
 \sqrt{3}  i & -  i & 0 & 0 & 0 & 0 & 0 \\
 0 & 0 & -  i & -  i & - \mathrm i & 0 & 0 
   \end{pmatrix} \ ,  \\
  & \lambda_z = \begin{pmatrix}
   0 & 0 & 0 & 0 & 0 & 0 & 0  \\
   0 & 0 & - 2  i & - 2  i & - 2  i & 0 & 0 \\
   0 & 2  i & 0 & 0 & 0 & 0 & 0 \\
  0 & 2  i & 0 & 0 & 0 & 0 & 0 \\
  0 & 2  i & 0 & 0 & 0 & 0 & 0 \\
  0 & 0 & 0 & 0 & 0 & 0 & -  i \\
  0 & 0 & 0 & 0 & 0 &  i & 0
   \end{pmatrix}
   \end{split} \ .
\end{align} 
Here, $\mathds{1}_m$ is the $m \times m$ unity matrix.

The distribution funtion $f_{\vec{k}}$ is calculated by solving the semiclassical Boltzmann equation for a stationary and spatially homogeneous system,
\begin{align}\label{eq:Boltzmann}
    \dot{\vec k} \frac{\partial f_{\vec k}}{\partial \vec{k}} = \left( \frac{\partial f_{\vec{k}}}{\partial t} \right)_\text{scatt} \ .
\end{align}
The left-hand side corresponds to the influence of external fields on the distribution function. In the presence of an external electric field $\vec E$, it is given by the semiclassical equation of motion 
\begin{align}\label{eq:kdot}
    \dot{\vec{k}}= - \frac{e}{\hbar} \vec E
\end{align}
with $e$ the absolute value of the elementary charge. The right-hand side of Eq.~\eqref{eq:Boltzmann} corresponds to the scattering term. Using the constant relaxation time approximation
\begin{align}\label{eq:scattering_term}
 \left( \frac{\partial f_{\vec{k}}}{\partial t} \right)_\text{scatt} = - \frac{1}{\tau_0} \left(f_{\vec{k}} - f_{\vec k} ^0 \right)   
\end{align}
with $\tau_0$ the constant momentum relaxation time (set to $\tau_0=1 \si{\pico\second}$ in our calculations) and $f_{\vec{k}}^0$ the equilibrium distribution function (which is the Fermi Dirac distribution function for fermions), the Boltzmann equation~\eqref{eq:Boltzmann} is solved by
\begin{align}\label{eq:Boltzmann_solution}
    f_{\vec k} = f_{\vec k}^0 + \frac{\partial f _{\vec k} ^0 }{\partial \epsilon} e \tau_0 \vec v_{\vec k} \cdot \vec E \ .
\end{align}
The group velocity $\vec v_{\vec{k}}$ is the expectation value of the velocity operator $\hat{\vec{v}}$,
\begin{align}\label{eq:velocity}
\vec v_{\vec{k}}= \Braket{\Psi_{\vec{k}} | \hat{\vec{v}}| \Psi_{\vec{k}}}
\end{align}
with $\hat{\vec{v}}= \nicefrac{i}{\hbar} \left[ H, \hat{\vec{r}}\right] = \nicefrac{1}{\hbar} \nicefrac{\partial H}{\partial \vec{k}}$ and $\hat{\vec{r}}= i \nicefrac{\partial }{\partial \vec{k}} $.

\section*{Acknowledgements}

This project received funding from the ERC Advanced Grant “FRESCO” ($\#$833973), the QuantERA project "QUANTOX" (ANR-18-QUAN-0014), the French ANR projects "QUANTOP" (ANR-19-CE47-0006-01) and "CORNFLAKE"
(ANR-18-CE24-0015-01), and Intel’s Science Technology Center – FEINMAN. M.B. thanks the Alexander von Humboldt Foundation for supporting his stays at Martin-Luther-Universität Halle. The authors thank J.-P. Attané, L. Vila, C. Proust and D. Vignolles for useful discussions.

\section*{Author contributions statement}

M.B. proposed the study and led it with I.M. L.M.V.A. prepared the samples with help from J.R. and P.L. S.V. led the ARPES study with S.M. and L.M.V.A., with support from J.B., R.S., J.R., F.B. and P.L. S.V. analysed the ARPES data and discussed them with A.J., B.G., L.M.V.A., S.M., J.B., N.B., I.M. and M.B. B.G. performed the TB fits with help from A.J. and I.M.. A.J. performed the EE calculations with help from B.G. and I.M. M.B., S.V., A.J. and B.G. wrote the paper with inputs from all authors.

\section*{Additional information}

The authors declare no competing interests.

\end{document}